\documentclass{appolb}
\usepackage{graphicx}
\begin{document}
\title{Discrimination of effective radiative and collisional in-medium energy-loss models by their effects on angular jet structure
\thanks{Presented at the XXIV Cracow Epiphany Conference, IFJ-PAN Cracow, 9.1.-12.1.2018.}%
}
\author{Martin Rohrmoser,
\address{Institute of Physics, Jan Kochanowski University, 25-406 Kielce, Poland, mrohrmoser@ujk.edu.pl}
\\
{Pol-Bernard Gossiaux, Thierry Gousset, J\"org Aichelin}
\address{SUBATECH, Rue 4 Alfred Kastler, 44307 Nantes, France}
}
\maketitle
\begin{abstract}
Energy-loss studies of hard particle probes produced in heavy ion collisions have often been used to get information on the interactions within the medium of a quark-gluon plasma (QGP).
However, with the study of in-medium energy-loss of individual particles alone, it remains still ambiguous, whether the occured decrease in particle energy is caused by predominantly radiative or collisional energy-loss mechanisms.
Focusing on the in-medium energy-loss of hard jet-partons, we propose additional studies of the angular jet-structure as a means to further constrain the energy-loss mechanisms.
\end{abstract}
%\PACS{PACS numbers come here}
  
\section{Introduction}
Highly energetic and/or heavy particles (hard probes) represent an excellent tool to study the quark-gluon plasma (QGP), as they are likely created in the initial stages of the collisions and do not attain thermal equilibrium.
Traversing the QGP, the hard probes in general lose energy, due to interactions with the medium. These energy-loss mechanisms are mostly described theoretically by either collisional or radiative interactions with the medium, as well as combinations thereof.
The combination of the observables of the nuclear modification factor $R_{\rm AA}$ and the elliptic flow $v_2$ allows to put constraints on these processes.
However, it remains still ambiguous, how much collisional and radiative processes contribute to the total energy loss, since theoretical models that contain either radiative processes as well as models that use a combination of both radiative and collisional processes allow to describe $R_{\rm AA}$ and $v_2$ equally well (cf. e.g.~\cite{Andronic:2015wma} for a current review).

$R_{\rm AA}$ and $v_2$ are observables corresponding to the distributions of individual hard probes.
However, radiative interactions with the medium may act as an additional source for correlated particle pairs, while purely collisional processes will not. 
Thus, observables for the multiple hard particles contained within jets may provide additional constraints on parton energy-loss mechanisms. 
An example are two-particle correlations, which have been studied intensively at experiments at the LHC. 
Recently, the values of the jet shape parameter $\rho$ that allow to gain insight on the angular structure of jets have been obtained at the CMS experiment~\cite{Sirunyan:2018jqr}. 

In these proceedings we argue that radiative and collisional energy loss mechanisms yield qualitatively different contributions to the jet shapes. 
To this end, three effective model approaches were used to describe collisional as well as radiative jet-medium interactions. 
While these approaches are rather simplistic, they provide an overall consistent framework for the interactions with the medium and are, thus, particularly well suited for comparisons between different types of energy-loss mechanisms. 

Numerical results for these effective models were obtained from of a Monte-Carlo algorithm. The medium effects on jet-evolution are implemented in parallel to collinear parton splitting due to bremsstrahlung, which is already present in the vacuum. The jet-evolution due to bremsstrahlung in the algorithm represents a Monte-Carlo simulation of the Dokshitzer-Gribov-Lipatov-Altarelli-Parisi (DGLAP) evolution of jets with leading order splitting functions, starting from an initial parton. 

\section{Effective Models for in-medium jet-evolution}

In general, we focus on a description of timelike parton cascades, generated by an initial quark with a maximal virtuality of $Q_\uparrow$ and energy $E_{\rm ini}$.
The partons in the cascades undergo multiple collinear splittings, until their virtualities $Q$ reach a minimal scale $Q_\downarrow$. 

In the vacuum, jet evolution is described via multiple bremsstrahlung emissions, which follows the DGLAP equations. Numerical results have been obtained in a Monte-Carlo simulation of the corresponding jets with splitting functions and the Sudakov factors of leading order (LO) in perturbative QCD. 

The effects on jet particles of multiple scatterings with medium particles were summarized in a continuous change of the cascade-parton four-momenta during their in-medium propagation.
It was assumed that the jet-medium interactions yield only a small perturbation to the vacuum-jet evolution.
Thus, the parton splittings were selected by means of the LO-Sudakov factors and LO-splitting functions that had already been used for the DGLAP-evolution of timelike parton cascades in the vacuum.
The four-momenta of the produced intermediate partons were then evolved in the medium following the respective approach to jet-medium interactions (which is described in detail further below) over the parton life-time $\tau$. 
In the parton rest-frame the life-time can be estimated to be of the order of $1/Q$, which yields in the lab-frame
\begin{equation}
\tau=\frac{E}{Q^2}\,,
\label{eq:timeest}
\end{equation}
where $E$ is the parton energy. 
Almost identical approaches have been used before~\cite{Zapp:2008gi,Renk:2008pp}.
Since both $E$ and $Q$ can be affected by jet-medium interactions, Eq.~(\ref{eq:timeest}) has to be solved self-consistently (as noted in~\cite{Renk:2008pp}). 
The parton four-momenta obtained from the self-consistent solution are used in the selection of a new parton splitting, if $Q>Q_\downarrow$.
If $Q<Q_\downarrow$ the particle is not subjected to any further interactions or splittings.

The approach to radiative energy loss is largely based on an early version of YAJEM~\cite{Renk:2008pp}, as it was assumed that medium induced radiation can be simulated by a continuous increase of $\hat{q}_R$ over time $t$ in the squared parton virtuality.
For quarks, the increase in $Q^2$ is described by
\begin{equation}
\frac{d }{dt}Q^2=\hat{q}_R\,.
\label{eq:yajemcont}
\end{equation}
For gluons, the right hand side of Eq.~(\ref{eq:yajemcont}) is multiplied by a factor 
$C_A/C_F$.
The increase in virtuality leads to an additional amount of radiated jet particles, and corresponds to an on average shorter life time of the intermediate particles, and vice versa. 
Thus, we argue that the model is well suited to describe the qualitative effects of medium induced radiation. 

The change in $Q^2$ in Eq.~(\ref{eq:yajemcont}) requires that at least one four-momentum component changes as well. 
We choose $\dot{E}^2=\dot{Q}^2=\hat{q}_R$, since this is the only choice that leaves the parton three-momenta $\vec{p}$ completely invariant, and, thus, allows to simulate purely collisional effects.
Physically, this choice corresponds to an energy transfer from the medium to the jet.
However, due to additional parton-splittings, the final particles in this medium model have smaller energies than in vacuum-cascades.

For an effective description of collisional energy loss, a transverse momentum transfer $\hat{q}_C$ and a longitudinal drag-force $\vec{A}$ are used, i.e.
\begin{eqnarray}
\hat{q}_C(t):=\frac{d \langle\vec{p}_\perp\rangle^2}{dt}\,,
&&
\vec{A}(t):=-\frac{d}{dt}\langle \vec{p}_L \rangle\,,
\end{eqnarray}
where $\vec{p}_L$ is the three-momentum component in direction of the incident cascade particle and $\vec{p}_\perp$ is the component orthogonal to $\vec{p}_L$.
The medium is assumed to be locally in thermal equilibrium, when the jet-medium interactions occur.
Thus, $\hat{q}_C$ and $\vec{A}$ are related by an Einstein-Smoluchowski relation.
In particular, from Ref.~\cite{Berrehrah:2014kba}, 
\begin{equation}
\frac{\hat{q}_C}{A}\approx 0.56+1.44\frac{T}{T_c}\,,
\label{eq:qhatAratio}
\end{equation}
was obtained with the critical temperature $T_c=0.158$~GeV. 
Following results of the jet collaboration~\cite{Burke:2013yra} a proportionality $\hat{q}_C\propto T^3$, more specifically for this work $\hat{q}_C=7T^3$, was assumed.

For the numerical implementation of the medium models, time was discretized into small steps $\Delta t$. Then, a direction $\vec{n}:= \vec{p}_\perp/\|\vec{p}_\perp\|$ for the action of the transverse kicks was determined by selecting an azimuthal angle in the plane orthogonal to $\vec{p}_L=\vec{p}(t)$. Per timestep $\Delta t$, the three-momenta and the virtuality change in the following way
\begin{eqnarray}
&&Q(t)\mapsto Q(t+\Delta t)=\sqrt{Q(t)^2+r\Delta t \hat{q}_R(t)}\,,\nonumber\\
&&\vec{p}(t)\mapsto\vec{p}(t+\Delta t)=\nonumber\\&&\vec{p}(t)+s\left(\vec{n}(t)\sqrt{\hat{q}_C(t)\Delta t}-A(t)\Delta t \frac{\vec{p}(t)}{\|\vec{p}(t)\|}\right)\,,
\label{eq:hybdpdqD}
\end{eqnarray}
where the parameters $r$ and $s$ specify the effective model of jet-medium interaction: 
the purely radiative model A ($r=1$, $s=0$), the purely collisional model B ($r=0$, $s=1$), as well as a hybrid model C ($r=1$, $s=1$).

The changes per $\Delta t$ in Eq.~(\ref{eq:hybdpdqD}) correspond to the following changes in parton-energy $E$
\begin{eqnarray}
&&E(t)\mapsto E(t+\Delta t)=\left(E(t)^2+\right.\nonumber\\&&\left.\Delta t(r\hat{q}_R(t)+s(\hat{q}_C(t)-2\|\vec{p}(t)\|A(t)))+\mathcal{O}\left(\Delta t^2\right)\right)^{\frac{1}{2}}\!.
\label{eq:yajemDEincr}
\end{eqnarray}
For models that contain collisional energy loss, the parton energy decreases for parton momenta $\|\vec{p}\|\gg T$, but increases if
\begin{equation}
\|\vec{p}\|<\frac{\hat{q}_C+r\hat{q}_R}{2A}\,.
\label{eq:ptcomparD}
\end{equation}

For simplicity we assumed $\hat{q}:=\hat{q}_C=\hat{q}_R$ and used the fit 
\begin{equation}
\hat{q}(t)=\frac{a}{(b+t)^c}\,.
\label{eq:qhatdev}
\end{equation}
from Ref.~\cite{Renk:2008pp} with $b=1.5$~fm/c and $c=2.2$. The parameter $a$ is determined by the overall transfer 
\begin{equation}
\Delta Q^2:=\int_{t_0}^{t_f}\hat{q}(t)dt\,,
\label{eq:dQ2def}
\end{equation}
where $t_0=0$ and $t_f=L=10$~fm/c was assumed, which yields $a\approx\frac{\Delta Q^2}{0.47}$. 

\section{Jet shapes}
An observable~\cite{Sirunyan:2018jqr} that has recently gained some experimental interest are the so called jet shapes $\rho(\Delta r)$, defined as
\begin{equation}
\rho(\Delta r):=\frac{1}{\delta r}\frac{\sum_{\rm jets}\sum_{{\rm tracks}\in (r_a,r_b)}p_T^{\rm trk}}{\sum_{\rm jets}\sum_{{\rm tracks}}p_T^{\rm trk}}\,,
\label{eq:defrho}
\end{equation}
where the $p_T^{\rm trk}$ are transverse momenta of jet particles. $\Delta r$ is defined as the radial distance $\Delta r:=\sqrt{\Delta \eta^2+\Delta \phi^2}$ with regard to the jet axis. In the numerator of Eq.~(\ref{eq:defrho}) only transverse momenta of particles are summed up, where the radial distance to the jet axis is inside the interval $(r_a,r_b)$ with $r_a=\Delta r-\delta r/2$ and $r_b=\Delta r+\delta r/2$. In the denominator the transverse momenta of all jet-particles are summed up.

By definition jet shapes are infrared and collinearly safe observables. 
This property is important for making comparisons with the results from the Monte-Carlo algorithm that was discussed in the previous section: So far, the Monte-Carlo algorithm for partonic cascades has not yet been convoluted with a hadronization mechanism. Instead the cascade evolution is cut off at the scale $Q_\downarrow$. Thus, the jet shapes $\rho$ are expected to depend only weakly on the value of $Q_\downarrow$, or on particular choices of hadronization mechanisms.

\begin{figure}[htb]
\centerline{%
\includegraphics[width=10cm]{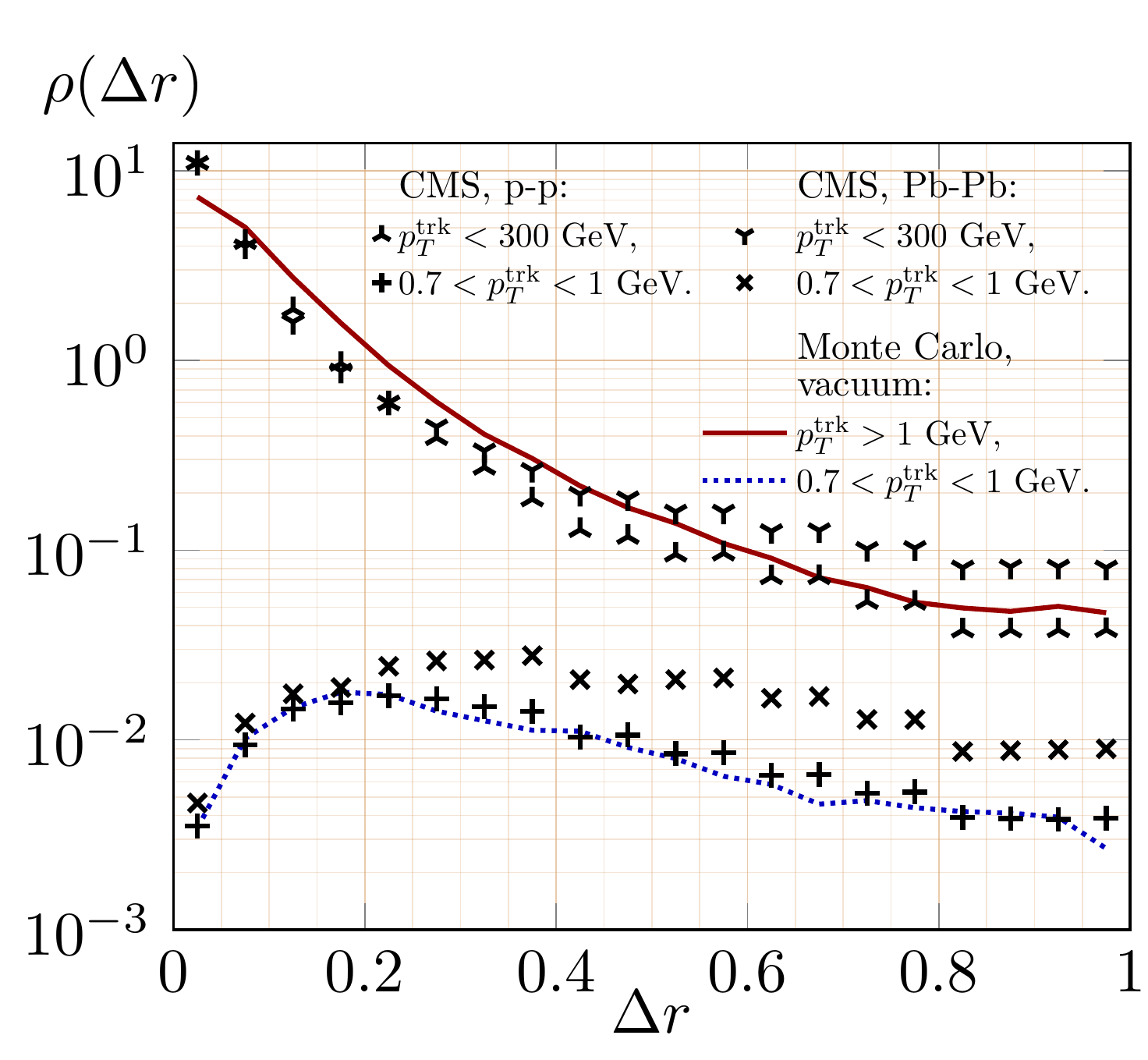}}
\caption{jet shapes from experimental data by CMS~\cite{Sirunyan:2018jqr} in p-p and central Pb-Pb collisions and for jets propagating in the vacuum, all of them together with their respective contributions from soft jet-particles.}
\label{fig:vacuum}
\end{figure}

\begin{figure}[h!!!]
\centering
\includegraphics[width=9.5cm]{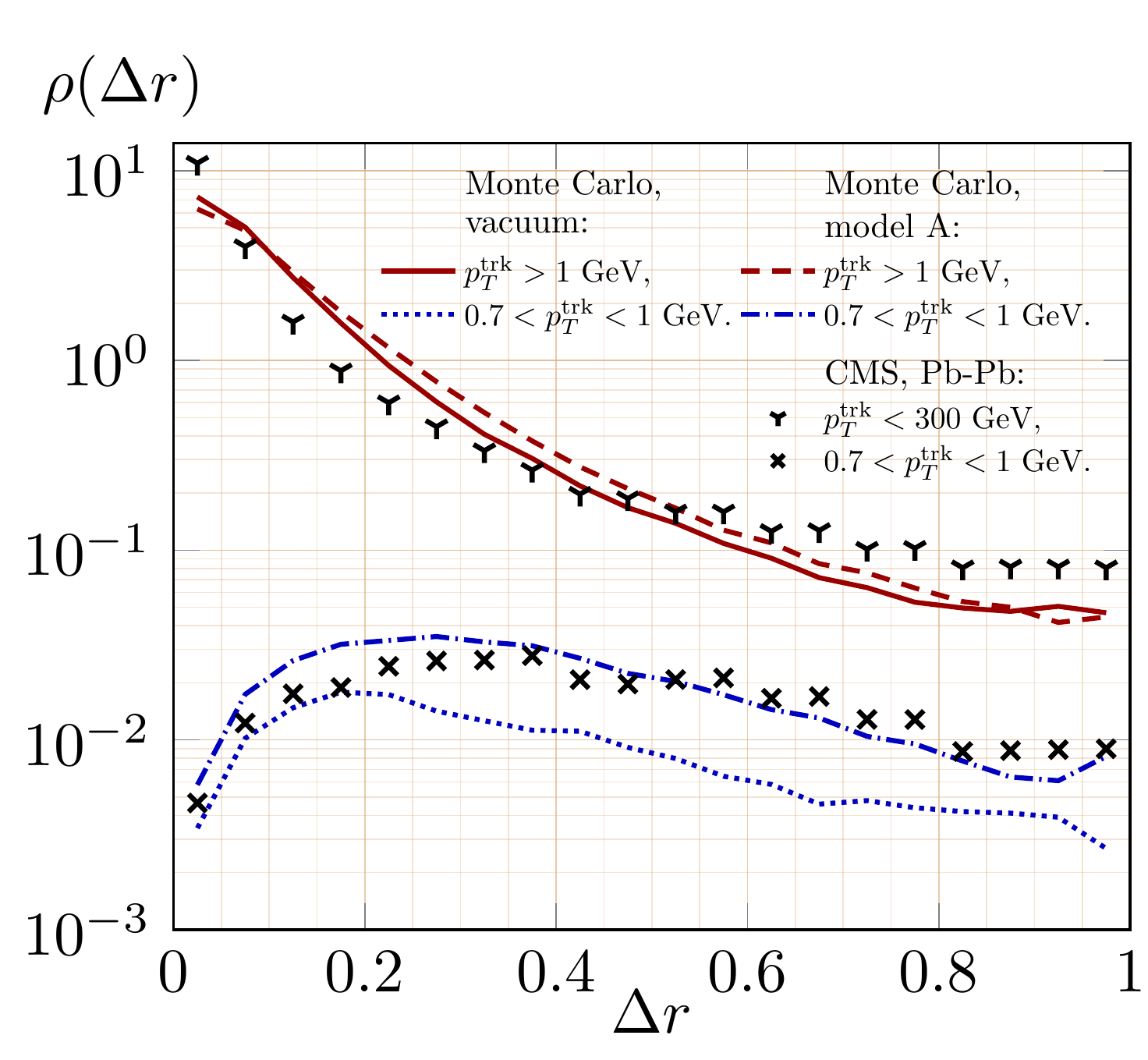}\\[-10mm]
\includegraphics[width=9.5cm]{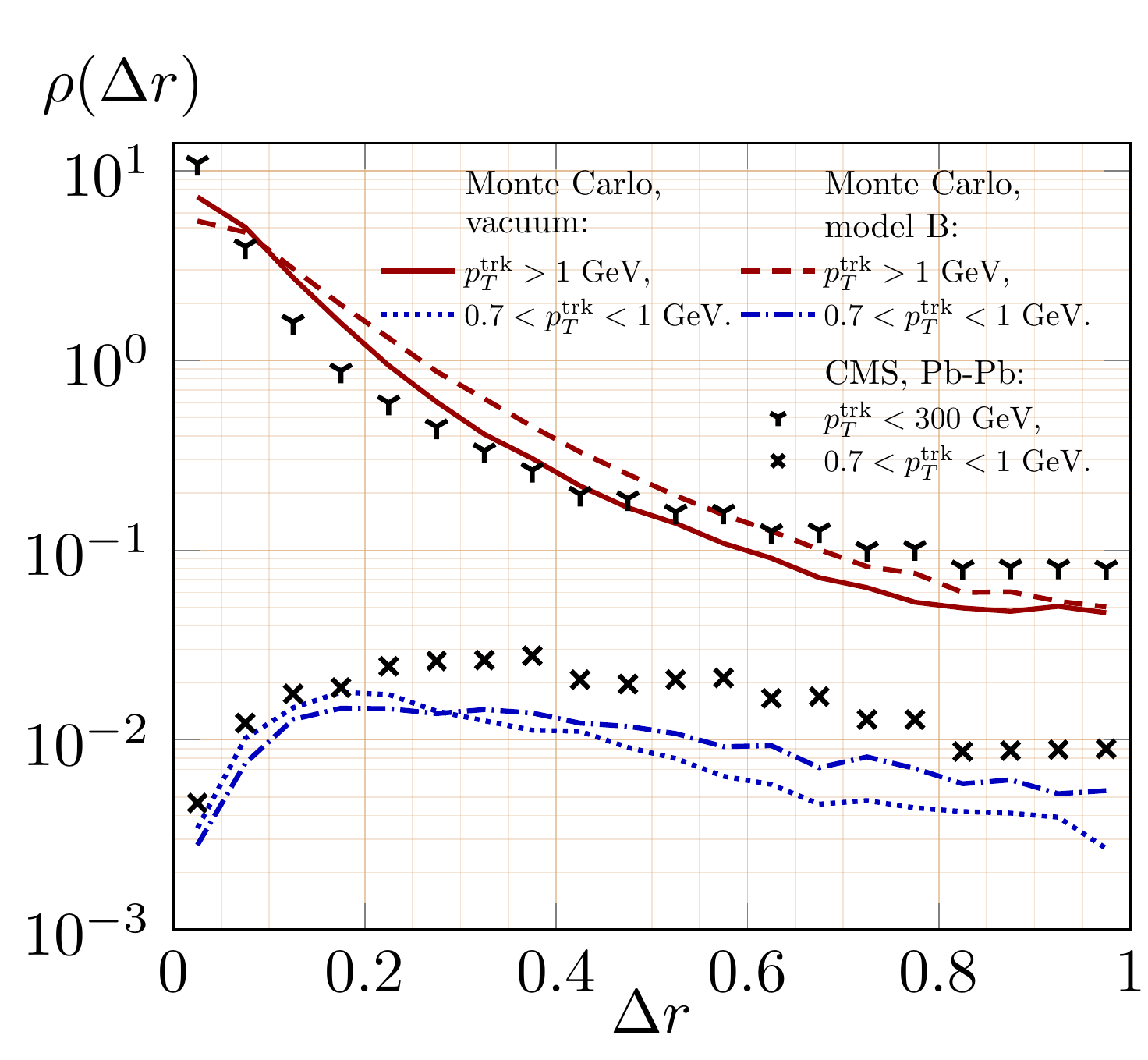}
%}
\caption{jet shapes from models A (upper panel) and B (lower panel) compared to experimental data by CMS~\cite{Sirunyan:2018jqr} in central Pb-Pb collisions and jets propagating in the vacuum, together with their respective contributions from soft jet-particles.}
\label{fig:A+B}
\end{figure}

Fig.~\ref{fig:vacuum} shows results for $\rho$ obtained from the Monte-Carlo simulation of parton cascades that evolve in the vacuum from an initial quark with $Q_\uparrow=E_{\rm ini}=200$~GeV down to a virtuality scale of $Q_\downarrow=0.6$~GeV together with experimental data from CMS for p-p collisions and Pb-Pb collisions of a centrality below $10\%$, both for $\sqrt{s}=5.02$~TeV. 
The Monte-Carlo results, just as the p-p collision data, show a strong decrease at small values of $\Delta r$ that becomes smaller at large $\Delta r$, although the decrease at small $\Delta r$ is less pronounced than in the experimental data. 
Fig.~\ref{fig:vacuum} also shows the contributions to $\rho(\Delta r)$ from soft jet particles, where $0.7<p_T^{\rm trk}<1$~GeV. 
For these soft contributions Monte-Carlo results and p-p collision data show a similar behavior.

The comparison between experimental data for p-p and Pb-Pb collisions in Fig.~\ref{fig:vacuum} reveals that the overall decrease of $\rho(\Delta r)$ is smaller in case of the heavy ion collisions, yielding also larger values of $\rho(\Delta r)$ at large $\Delta r$ values. This behavior corresponds to an increased radiation at large $\Delta r$ values, as well as the additional production of soft particles, where $0.7<p_T^{\rm trk}<1$~GeV (for additional experimental data, cf. Ref.~\cite{Sirunyan:2018jqr}).

Fig.~\ref{fig:A+B} shows the results for Monte-Carlo simulations of the purely radiative model A (upper panel) and the purely collisional model B (lower panel), both for $\Delta Q^2=3$~GeV$^2$, in comparison to experimental data from Pb-Pb collisions and the Monte-Carlo results for vacuum cascades.
While the Monte-Carlo results do not reproduce the experimental values, the results for all $p_T^{\rm trk}$ values from both models exhibit a smaller decrease with $\Delta r$ as compared to results from cascades in the vacuum.
For model A this broadening effect is rather small (as compared to the behavior for model B). However, the soft contributions are largely increased for model A - in agreement with data from Pb-Pb collisions at large $\Delta r$, but corresponding to values for $\rho$ at small $\Delta r$ that are too large. 
We conclude from the behavior of the $\rho(\Delta r)$ results for model A that the broadening effects in this purely radiative model are mainly due to the medium-induced radiation of very soft particles.
For the purely collisional model B the broadening effects of the medium are even larger than for model A, while the contributions from soft particles are only mildly affected by the medium. In contrast to the behavior of the results for model A, the soft contributions for model B are not enhanced at all $\Delta r$ values, but only at large $\Delta r$.
We argue that the behavior of the results for model B can be explained by properties of the stochastic transverse forces and the drag force: These forces affect hard as well as soft particles (much in contrast to medium induced radiation, which predominantly leads to the production of soft particles). Due to the deflections by the effective forces, the angles of particle three-momenta, represented by the corresponding $\Delta r$ values, are larger in the medium than in the vacuum.

\begin{figure}[htb]
\centerline{%
\includegraphics[width=11.5cm]{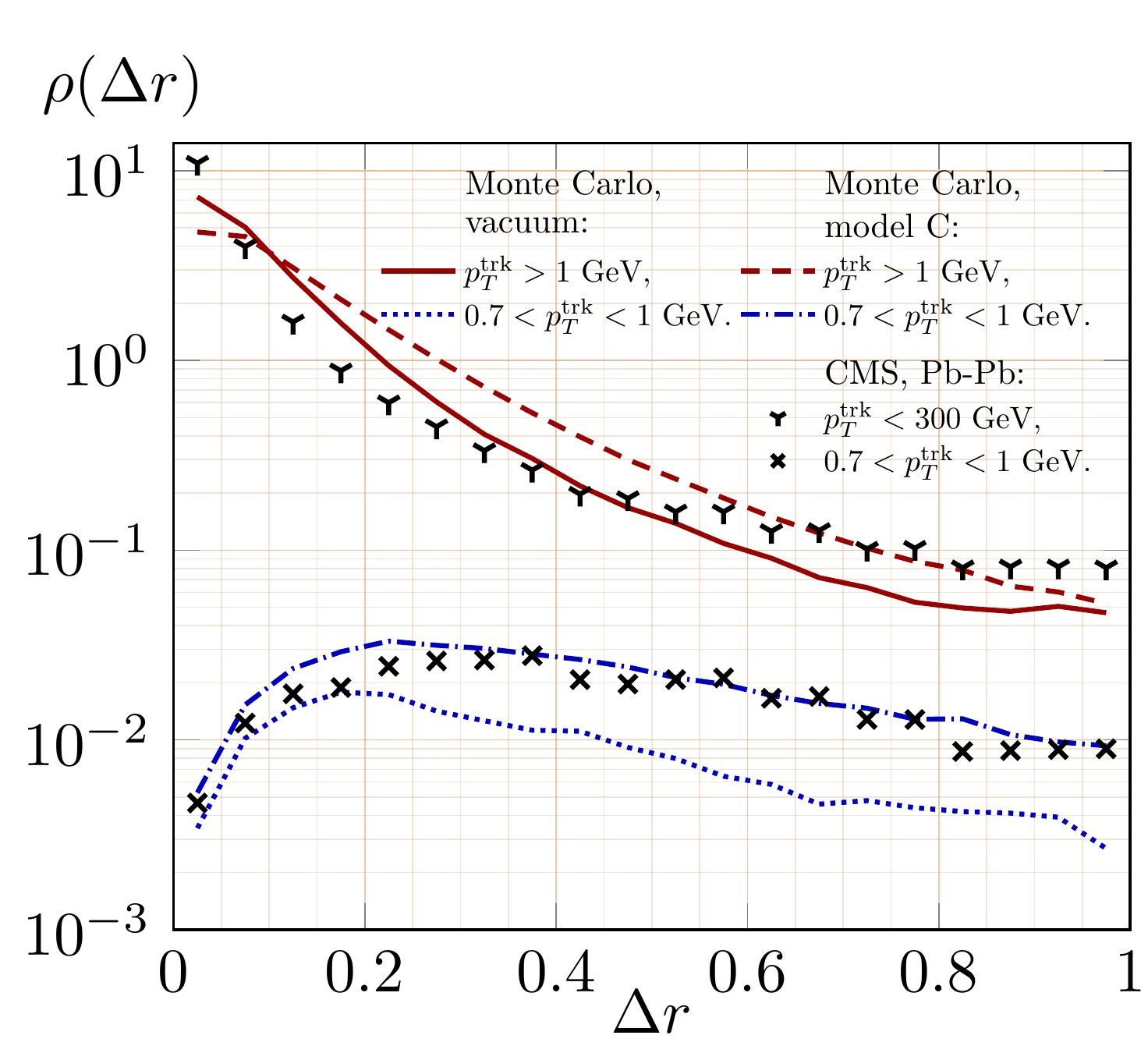}
}
\caption{Same as Fig.~\ref{fig:A+B} but for the hybrid model C.}
\label{fig:C}
\end{figure}

Finally, Fig.~\ref{fig:C} shows the results for the hybrid model C, also for $\Delta Q^2=3$~GeV$^2$: As expected, this model exhibits the largest broadening effects, corresponding to an enhancement specifically at large angles, due to collisional effects, as well as a largely increased production of soft particles due to radiative effects.

\section{Summary}
We have presented a simplistic, yet consistent framework of effective models for jet-medium interactions. 
Three models, a purely radiative model A, a purely collisional model B, and a hybrid model C were implemented in a Monte-Carlo algorithm for the simulation of jets in the medium, in order to study medium effects on the jet-shape observables.

While the models, and the algorithms implemented neglect many details and are therefore not suitable to reproduce experimental data quantitatively, the qualitative behavior of jet-shape data from, e.g. the measurements at CMS can be obtained.
Radiative energy-loss mechanisms were found to lead to a higher production of soft particles, while collisional energy-loss mechanisms deflect particle three-momenta to larger angles $\Delta r$ with regard to the jet axis. 
The qualitatively different behaviors of the jet shapes for collisional and radiative energy loss, in particular with regard to their soft contributions, may be used as a tool to disentangle collisional from radiative in-medium energy-loss mechanisms.

\section{Acknowledgements}
This research was supported in part by the Polish National Science
Centre Grant No. 2015/19/B/ST2/00937.


\begin{thebibliography}{1}
\bibitem{Andronic:2015wma}
  A.~Andronic {\it et al.},
  %``Heavy-flavour and quarkonium production in the LHC era: from proton–proton to heavy-ion collisions,''
  Eur.\ Phys.\ J.\ C {\bf 76} (2016) no.3,  107
  doi:10.1140/epjc/s10052-015-3819-5
  [arXiv:1506.03981 [nucl-ex]].

\bibitem{Sirunyan:2018jqr}
  A.~M.~Sirunyan {\it et al.} [CMS Collaboration],
  %``Jet properties in PbPb and pp collisions at $\sqrt{s_\mathrm{NN}} =$ 5.02 TeV,''
  arXiv:1803.00042 [nucl-ex]. 
  
\bibitem{Zapp:2008gi}
  K.~Zapp, G.~Ingelman, J.~Rathsman, J.~Stachel and U.~A.~Wiedemann,
  %``A Monte Carlo Model for 'Jet Quenching',''
  Eur.\ Phys.\ J.\ C {\bf 60} (2009) 617
  doi:10.1140/epjc/s10052-009-0941-2
  [arXiv:0804.3568 [hep-ph]].
  
\bibitem{Renk:2008pp}
  T.~Renk,
  %``Parton shower evolution in a 3-d hydrodynamical medium,''
  Phys.\ Rev.\ C {\bf 78} (2008) 034908
  doi:10.1103/PhysRevC.78.034908
  [arXiv:0806.0305 [hep-ph]].
  
\bibitem{Berrehrah:2014kba}
  H.~Berrehrah, P.~B.~Gossiaux, J.~Aichelin, W.~Cassing and E.~Bratkovskaya,
  %``Dynamical collisional energy loss and transport properties of on- and off-shell heavy quarks in vacuum and in the Quark Gluon Plasma,''
  Phys.\ Rev.\ C {\bf 90} (2014) no.6,  064906
  doi:10.1103/PhysRevC.90.064906
  [arXiv:1405.3243 [hep-ph]].   

\bibitem{Burke:2013yra}
  K.~M.~Burke {\it et al.} [JET Collaboration],
  %``Extracting the jet transport coefficient from jet quenching in high-energy heavy-ion collisions,''
  Phys.\ Rev.\ C {\bf 90} (2014) no.1,  014909
  doi:10.1103/PhysRevC.90.014909
  [arXiv:1312.5003 [nucl-th]].
\end{thebibliography}
\end{document}